# Ultra-broadband Microwave Metamaterial Absorber

Fei Ding[1], Yanxia Cui[1,3], Xiaochen Ge[1], Feng Zhang[1], Yi Jin[1*], and Sailing He[1,2]

[1]*Centre for Optical and Electromagnetic Research, State Key Laboratory of Modern Optical Instrumentations, Zhejiang University, Hangzhou 310058, China*

[2]*Division of Electromagnetic Engineering, School of Electrical Engineering, Royal Institute of Technology, S-100 44 Stockholm, Sweden*

[3]*Department of Physics and Optoelectronics, Taiyuan University of Technology, Taiyuan 030024, China*

A microwave ultra-broadband polarization-independent metamaterial absorber is demonstrated. It is composed of a periodic array of metal-dielectric multilayered quadrangular frustum pyramids. These pyramids possess resonant absorption modes at multi-frequencies, of which the overlapping leads to the total absorption of the incident wave over an ultra-wide spectral band. The experimental absorption at normal incidence is above 90% in the frequency range of 7.8−14.7*GHz*, and the absorption is kept large when the incident angle is smaller than 60 degrees. The experimental results agree well with the numerical simulation.

*Corresponding author: jinyi@coer.zju.edu.cn

Electromagnetic metamaterials (MMs) have produced many exotic effects and devices, such as negative refraction [1−3], sub-wavelength imaging super-lens [4], and cloaking [5]. The novel properties of MMs come from the artificial atoms which can flexibly tailor the electric and magnetic response. MMs can be considered as an effective medium with complex permittivity ($\varepsilon$) and permeability ($\mu$) [6]. Most of the research in MMs has focused on the real parts of $\varepsilon$ and $\mu$, and typically it is desirable to reduce the loss represented by the imaginary parts of $\varepsilon$ and $\mu$[7]. However, the loss in MMs is also capable of creating useful phenomena and devices. Recently, Landy et al. have proposed a thin metamaterial absorber (MA), in which electric and magnetic resonance makes the absorber possess matched impedance to eliminate the reflection and strongly absorb the incident wave [8]. Since then, many MAs have been proposed and demonstrated from microwave to optical frequencies [9−14]. Because resonance is utilized in the process of absorption, the absorption bandwidth is often narrow, typically no larger than 10% with respect to the central frequency. In many cases, broadband absorption is required, such as solar energy harvesting. An effective method to extend the absorption band is to make the MA units resonate at several neighboring frequencies [11, 14]. Usually, the MAs are made of a patterned metal film above a homogeneous metal film with a dielectric spacer. Some efforts have been made to extend the absorption band of such MAs, but the absorption spectra are composed of discrete absorption peaks [11], or the absorption band is not wide enough [14]. Here, an ultra-broadband absorber will be demonstrated in the microwave band. This MA is composed of multilayered metallic-dielectric quadrangular frustum pyramids. Numerical simulation and experimental verification will be shown, and the explanation of the ultra-broadband absorption will also be given.

The schematic of the suggested MA is shown in Fig. 1(a), which consists of a periodic array of quadrangular frustum pyramids with a homogeneous metal film as the ground to block the transmission. A unit cell of the MA is shown in Fig. 1(b). A pyramid is composed of 20 metal patches with their width tapered linearly from the top to the bottom, and every two adjacent metal patches are separated by a dielectric patch. The metal is copper with the electric conductivity assumed to be $\sigma = 5 \times 10^7 s/m$. FR4 is used as the dielectric spacers with the relative permittivity and loss tangent equal to 4.4 and 0.02, respectively.

A numerical simulation is performed for the MA in Fig. 1(a) with a commercial program, CST Microwave Studio TM 2009 [16]. Periodic boundary conditions are used in the $x$ and $y$ directions, and a plane wave is incident downward on the MA with the electric field polarized along the $y$-direction (TE) as the excitation source. Transmission $T(\omega)$ and reflection $R(\omega)$ are obtained from the frequency-dependent S-parameter $S_{11}(\omega)$ and $S_{21}(\omega)$, that is, $T(\omega) = |S_{21}(\omega)|^2$ and $R(\omega) = |S_{11}(\omega)|^2$. The absorption is calculated as $A(\omega) = 1 - T(\omega) - R(\omega)$. Since the ground is of metal, $S_{21}(\omega)$ is nearly zero in the entire investigated frequency range. $A(\omega)$, $T(\omega)$ and $R(\omega)$ are shown in Fig. 2(a). There are 20 absorption peaks in the absorption spectrum with the absorptivity above 90%, corresponding to the number of dielectric patches between two neighboring metal interfaces. Relative absorption bandwidth (RAB) is used to describe the absorption performance of a MA, defined as $W_{RAB}=2(f_U-f_L)/(f_U+f_L)$ where $f_U$ and $f_L$ are the upper and lower limits of a frequency range with $A(\omega)$ above 90%, respectively. The MA in Fig. 1(a) possesses excellent absorption performance with $W_{RAB}$ equal to 53.1%, which is much larger than that reported in [11].

The fabricated example of the MA in Fig. 1(a) is shown in Fig. 1(c). The sample begins with a multilayered printed circuit board. Each copper layer with a thickness of 0.05*mm* is printed on a FR4 layer with a thickness of 0.15*mm*, and they form a composite layer. Twenty composite layers are then sandwiched using adhesive (whose dielectric constants are almost the same as FR4) with a thickness of 0.05*mm*, below which a copper film is added to suppress the transmission with a FR4 layer as the substrate. A MA is then fabricated using the standard mechanical milling method, which has an area of 200*mm*×200*mm*, including 256 units. The absorption performance of the test sample is verified by measuring the complex *S*-parameters in a microwave anechoic chamber. The experimental setup is the same as that in [17]. A R&S ZVB vector network analyzer is used, which works in the range of 1−15*GHz*. A broadband horn antenna focuses the microwave on the sample and another horn antenna serves as a receiver. The reflection is measured at an incident angle of 10 degrees, and the transmission is measured at normal incidence. The measured reflection is normalized with respect to a copper plane with the same dimensions as the sample, while the measured transmission is normalized with respect to the incident wave in free space. The measured transmission and reflection are then used to obtain the absorption. The experimental absorption is shown in Fig. 2(b) (see the solid red curve). One can see that the agreement between the experimental and simulated (blue curve) results is very good considering the tolerance in the fabrication and assembly. The RAB of the experimental absorption spectra is 61.6%, larger than that obtained by numerical simulation, which may be attributed to the fabrication error.

To better understand the physics of the suggested ultra-broadband MA, the electric and magnetic field distributions at some frequencies are depicted in Fig. 3. It is evident that, at a certain frequency, the electromagnetic field is resonantly localized and then absorbed at some part of the pyramids. At a smaller frequency, the electromagnetic field is localized at the bottom side of the pyramids, and as the frequency increases, the localized electromagnetic field moves gradually towards the top-side. Since neighboring metal patches vary slowly in width, the resonant electromagnetic field is not localized in the dielectric spacer between two neighboring metal patches, but diffuses into several dielectric spacers nearby. That is, several neighboring metal and dielectric patches together support a resonant mode. In such a resonant mode, electric and magnetic resonance both exist, and the resonant frequency is nearly inversely proportional to the patch width, which is similar to that in common MAs composed of an array of metal patches above a metal substrate with a dielectric spacing layer. The electric and magnetic resonance make the absorber impedance match well with the free space, and the incident wave is absorbed with low reflection. The pyramids of the suggested MA gradually increase in width from the top to the bottom and can support the resonance in a wide frequency band. The collection of the resonance at different frequencies results in the ultra-broadband absorption. At a certain specific frequency corresponding to the resonant mode supported by a pair of metal patches separated by a dielectric patch, the incident wave is most strongly absorbed and an absorption peak is formed on the absorption curve in Fig. 2 (a).

The ultra-broadband absorption can also be interpreted in another way. First, we investigate a periodic structure shown in Fig. 4(a), whose unit cell is composed of a copper patch on a FR4 patch. The width for the copper and FR4 patches is $w$, and the thickness is $t_d$ and $t_m$, respectively. The lattice constants of the periodic structure are $p_x=11mm$, $p_y=11mm$, and $p_z=t_d+t_m$ along the $x$, $y$, and $z$ directions, respectively. For $t_m=0.05mm$, $t_d=0.2mm$, and $w$ varying from $9mm$ to $5mm$, Fig. 4(b) shows the corresponding dispersion curves when the Bloch mode propagates along the $z$ direction (the material loss is assumed to be zero). In Fig. 4(b), one can see that when $w=9mm$, the cutoff frequency is about $f_c=8$ $GHz$, and when the frequency is near $f_c$, the Bloch mode is a slow wave. When $w=5mm$, then $f_c$ increases to about $14.3GHz$. Now, consider the pyramidal MA structure in Fig. 1(a), where the width of the pyramids is tapered linearly from the top ($W_t=5mm$) to the bottom ($W_f=9mm$). The

MA can dramatically slow down and trap an incident wave of frequency $f$ at some position where $f_c$ is equal to $f$. Due to the practical loss of copper and FR4, the incident wave is then strongly absorbed. $f_c$ varies in a broad band of $8-14 GHz$ for different position along the height of the pyramids, thus one can get an ultra-broadband MA.

After the mechanism of ultra-broadband absorption has been explained, the influence of some structure parameters on the absorption can be easily understood. As discussed above, the incident wave at a large frequency is mainly absorbed on the top parts of the pyramidal cells. Thus, with the other parameters fixed, the top width of a pyramidal cell may nearly determine the upper frequency limit of RAB. Fig. 5(a) shows the corresponding simulated absorption spectra when $W_t$ is equal to $4mm$, $5mm$, and $6mm$, respectively. One can see that as $W_t$ is reduced, $f_U$ becomes larger while $f_L$ is nearly fixed, which results in an increase of the $W_{RAB}$. However, $W_t$ cannot be too small, otherwise neighboring metal patches in a unit will vary too quickly in width and the frequency difference between their supported resonant modes becomes too large and some strong oscillation will appear on the absorption spectra [see the blue curve in Fig. 5(a)]. To eliminate such oscillation, when the height of pyramids is fixed, more metal and dielectric patches need to be added with $t_m/t_d$ fixed. As an example, when $W_t$ is equal to $5mm$, and the number of composite layers is changed from 20 to 30 and 40, the corresponding absorption spectra are shown in Fig. 5(b). The oscillation becomes weaker and nearly disappears, leaving a smooth absorption spectrum.

At last, the polarization and angular dependence of the MA is discussed. Due to the symmetry of the designed MA, the absorption is independent of polarization. And the absorption effect is robust for non-normal incidence. For the sake of convenience, TE polarization is taken as an example. Simulation and experiment are performed to verify the angular dispersion of the absorption at various incident angles for TE configuration, which is shown in Fig. 5(c) and Fig. 5(d). Experimental results agree well with the simulation results, and the disagreement stems from the broadening of the experimental absorption spectrum, which originates from the fabrication error of the sample and the background testing signal. The absorption is nearly independent of the incident angle below 40 degrees, and as the incident angle increases further, the absorption becomes weaker. Nevertheless, the absorption still remains above 80% even when the incident angle reaches 60 degrees.

In conclusion, a microwave ultra-broadband MA has been proposed, fabricated and characterized, which achieves nearly unity absorption from 8GHz to 14GHz over a wide angular range. The absorption band can easily be tuned by changing the width grade of the adrangular frustum pyramids. Importantly, the design idea has the ability to be extended to other frequencies, like terahertz, infrared and optical frequencies. Such a feature may open a door to a controllable ultra-broadband absorber or thermal emitter.

We acknowledge Dr. Hongsheng Chen from the Electromagnetic Academy at Zhejiang University for his help in the sample test. F. Ding would thank Yingran He for useful discussion. This work is partially supported by the National Natural Science Foundation of China (Nos. 60990322, 60901039 and 61178062).

**Figures and Figure Captions:**

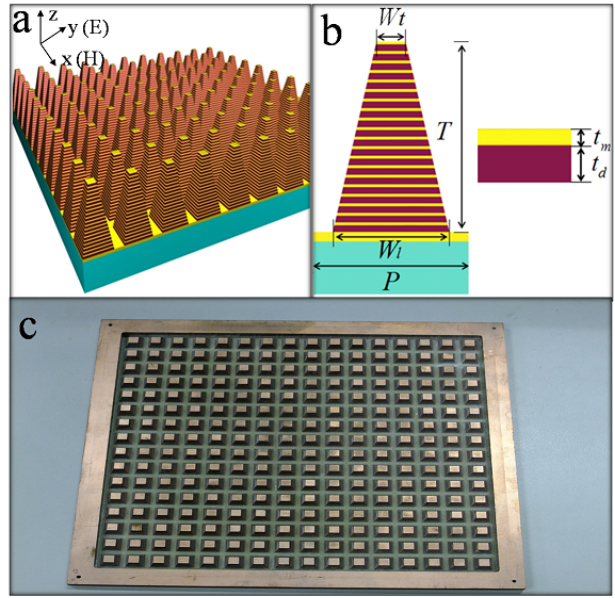

Fig. 1. (Color online) Design and fabrication of a microwave ultra-broadband MA. (a) Three-dimensional illustration of the simulated MA, (b) schematic of a MA unit cell, and (c) photograph of the fabricated sample. The optimized dimensions of a unit are $W_t$=5mm, $W_l$=9mm, $P$=11mm, $t_m$=0.05mm, $t_d$=0.2mm, and $T$=5mm. Subscript "m" represents copper, and "d" for FR4.

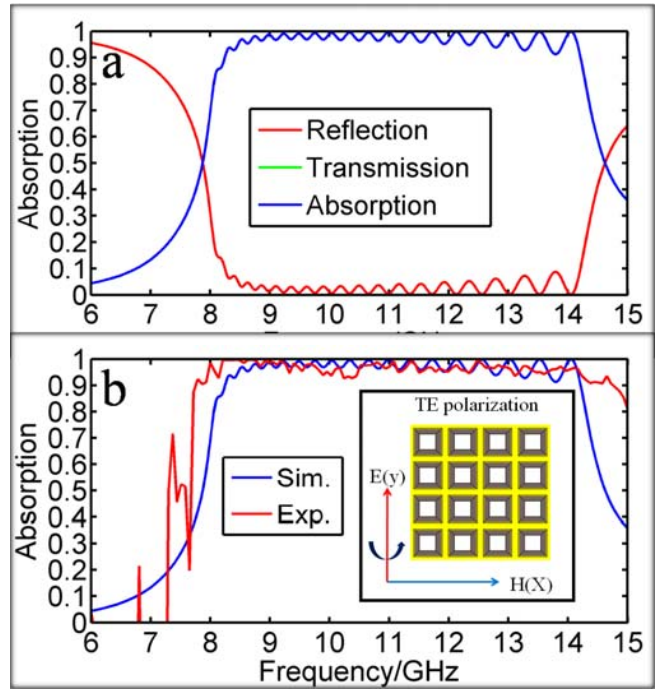

Fig 2. (Color online) Simulated and experimental absorption performance of the MA in Fig. 1. (a) Numerical simulation. Reflection (red line), transmission (green line), and absorption (blue line). (b) Comparison between the experimental absorption (red line) and simulated absorption (blue line). Inset in (b) shows the configuration of the incident wave.

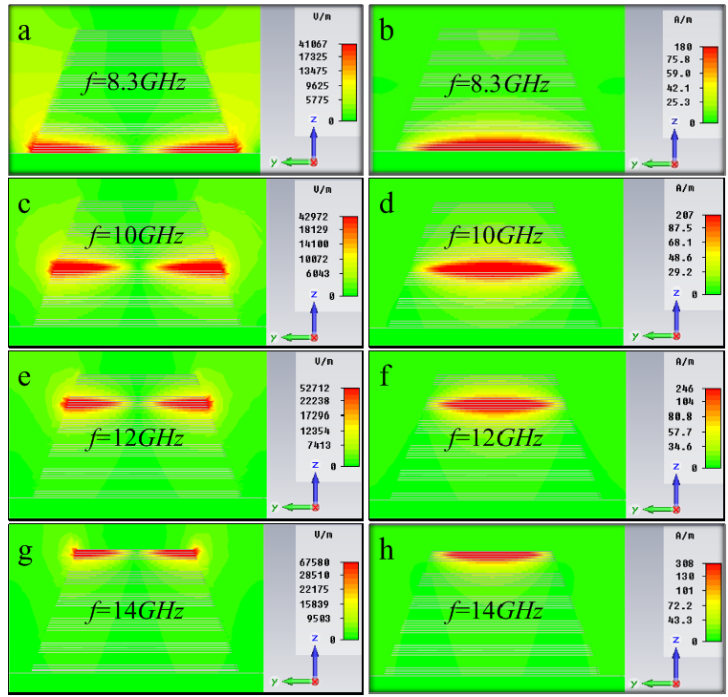

Fig. 3. (Color online) Simulated electric and magnetic amplitude distributions on the central cross section of a unit cell at some frequencies. (a), (c), (e), and (g) are for the electric amplitude at 8.3*GHz*, 10*GHz*, 12*GHz*, 14*GHz*, respectively. (b), (d), (f), and (h) are for the magnetic amplitude at 8.3*GHz*, 10*GHz*, 12*GHz*, 14*GHz*, respectively.

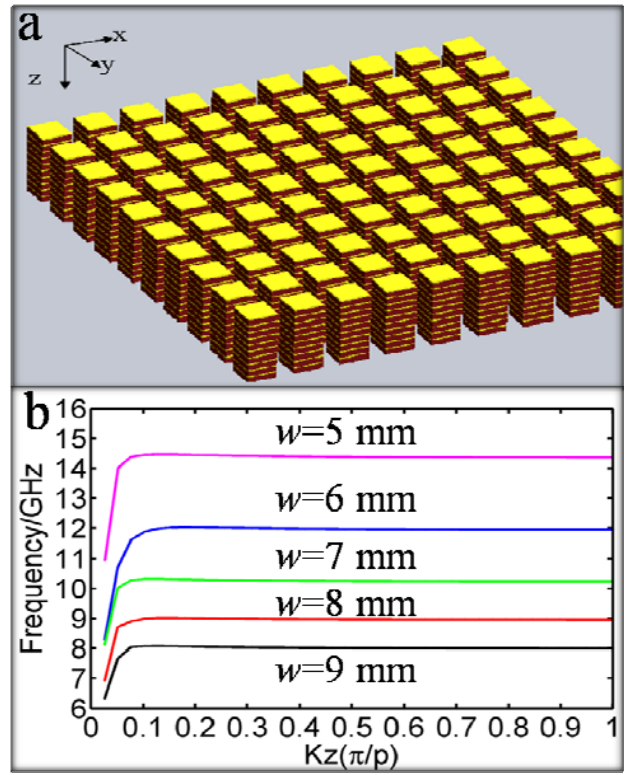

Fig. 4. (Color online) Mechanism of ultra-broadband absorption. (a) A periodic metal-dielectric structure. (b) Dispersion curves of the structure in (a) for $p_x=p_y=11mm$ and $p_z=0.25mm$ with different width, $w=5mm$ (magenta line), $w=6mm$ (blue line), $w=7mm$ (green line), $w=8mm$ (red line), $w=9mm$ (black line). Only the lowest energy band is shown.

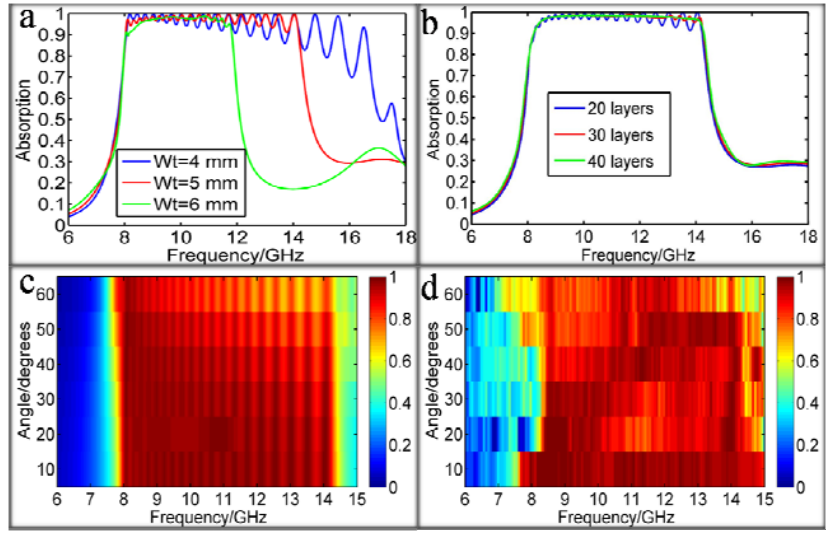

Fig. 5. (Color online) Influence of some structure parameters on the absorption and angular dispersion of absorption. (a) Simulated absorption when the top width of the pyramids in Fig. 1(a) changes from 4*mm* to 6*mm* with the other parameters fixed. (b) Simulated absorption when the number of the composite layers of the pyramids in Fig. 1(a) changes with the other parameters fixed. Simulated (c) and experimental (d) angular absorption of the MA in Fig. 1 for TE configuration. The incident angle is varied from 10 degrees to 60 degrees in a step of 10 degrees in the experiment, and changed from 0 degree to 60 degrees in the simulation.